\documentclass[twocolumn,showpacs,preprintnumbers,bibnotes]{revtex4}
\usepackage{graphicx}
\usepackage{dcolumn}
\usepackage{bm}
\usepackage{amssymb}
\usepackage{amsmath}
\usepackage{epsfig}

\begin{document}

\title{Experimental demonstration of phase measurement precision beating
standard quantum limit by projection measurement}
\author{F. W. Sun$^{1}$}
\email{fwsun@mail.ustc.edu.cn}
\author{B. H. Liu$^{1}$}
\author{Y. X. Gong$^{1}$}
\author{Y. F. Huang$^{1}$}
\author{Z. Y. Ou$^{1,2}$}
\email{zou@iupui.edu}
\author{G. C. Guo$^{1}$}
\affiliation{$^{1}$Key Laboratory of Quantum Information,
University of Science and
Technology of China (CAS), Hefei 230026, People's Republic of China.\\
$^{2}$Department of Physics, Indiana University-Purdue University
Indianapolis \\
402 N. Blackford Street, Indianapolis, IN 46202 }
\date{\today }

\begin{abstract}
We propose and demonstrate experimentally a projection scheme to
measure the quantum phase with a precision beating the standard
quantum limit. The initial input state is a twin Fock state
$|N,N\rangle$ proposed by Holland and Burnett [Phys. Rev. Lett.
{\bf 71}, 1355 (1993)] but the phase information is extracted by a
quantum state projection measurement. The phase precision is about
$1.4/N$ for large photon number $N$, which approaches the
Heisenberg limit of $1/N$. Experimentally, we employ a four-photon
state from type-II parametric down-conversion and achieve a phase
uncertainty of $0.291\pm 0.001$ beating the standard quantum limit
of $1/\sqrt{N} = 1/2$ for four photons.
\end{abstract}

\pacs{42.50.Dv, 03.65.Ta, 07.60.Ly.}
\maketitle

The measurement of the magnitude of a physical quantity is one of
the main tasks of modern physics. The key question is what
precision can be achieved in the measurement. Principally, this is
governed by the laws of quantum mechanics. For example, the
precision of the quantum phase measurement can be intuitively
understood from the Heisenberg uncertainty principle for the phase
and photon number \cite{dirac} as
\begin{equation}
\Delta \phi \Delta N\geq 1\text{,}
\end{equation}%
where $\Delta \phi $ and $\Delta N$ are the fluctuation for the
phase and photon number. Therefore, the shot noise of $\Delta
N=\sqrt{\langle N\rangle}$ from a laser in coherent state gives
rise to the shot-noise limit or the so-called standard quantum
limit \cite{cav} for the phase measurement as $\Delta \phi \gtrsim
1/\sqrt{\left\langle N\right\rangle }$ with an average photon
number of $\left\langle N\right\rangle$.

However, the standard quantum limit is not the ultimate limit for
the precision in phase measurement. It has been surpassed with
squeezed state based interferometry \cite{bondurant,xiao}. The
ultimate limit was proven \cite{ou1,ou2} to be the Heisenberg
limit \cite{hei} of $1/\langle N\rangle$. A number of schemes were
proposed \cite{bondurant,holland,jacobson,ou2,bollinger,campos}
that can reach this limit. But none of them was realized
experimentally, primarily due to the fragile effect of loss in the
system.

Recently, research focus is shifted to the maximally entangled
photon-number state (MES) or the so-called NOON state
\cite{bollinger,ou2,pan,mitchell,bow} in the form $(\left\vert
N,0\right\rangle +\left\vert 0,N\right\rangle )/\sqrt{2}$, which,
due to a multi-photon interference effect, leads to a multi-photon
detection probability of
\begin{equation}
P_{MES}(\phi ) = (1+\cos N\phi )/2%
\text{.}\label{2}
\end{equation}
The advantage of using MES is that it is not as sensitive to loss
as the squeezed states. Loss will simply reduce the success
probability and increase the number of photons required for phase
measurement, but in a linear fashion. It has been demonstrated
that the above super-resolved phase dependence can be achieved
without the need of a NOON state. However, as recently pointed out
\cite{res,sun}, the super-resolved phase dependence in
Eq.(\ref{2}) does not necessarily lead to a phase measurement
precision exceeding the standard quantum limit.

In this letter, we will study the problem of precision phase
measurement with a different approach. We will combine the concept
of twin-photon state \cite{holland,campos} with recently developed
quantum state projection method \cite{sun}. We find that the new
measurement scheme allows us to achieve a phase measurement
precision that is close to the Heisenberg limit. We demonstrate
the feasibility of the scheme with two-photon state and
four-photon state and obtain a phase measurement precision of
$0.501\pm 0.001$ and $0.291\pm0.001$ for these states,
respectively. These values are significantly better than
$1/\sqrt{2}$ and $1/2$ set by the standard quantum limit.

\begin{figure}[tbh]
\begin{center}
\includegraphics[width= 3in]{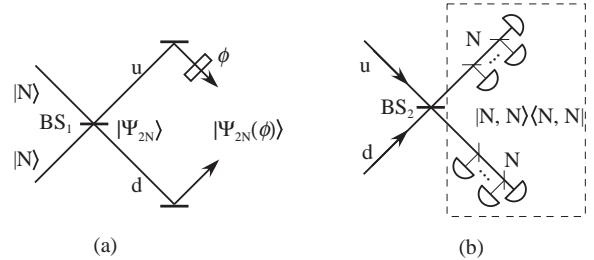}
\end{center}
\caption{Projection measurement for phase measurement with a
twin-Fock state. (a) Generation of the state $\left\vert \Psi
_{2N}(\phi )\right\rangle$ from a twin-Fock state; (b) Projection
measurement of $\mathbb{P}=|\Psi _{2N}\rangle\langle\Psi _{2N}|$.
The dashed box is a projection to $|N,N\rangle$. }
\end{figure}
\noindent {\it Theory} -- Consider a two-mode state that is
generated by injecting a twin Fock state $|N,N\rangle$ into a
50:50 beam splitter \cite{holland} as shown in Fig.1. The two-mode
state has the form of:
\begin{eqnarray}
\left\vert \Psi _{2N}\right\rangle &=&\sum_{k=0}^{N}(-1)^{N-k}\bigg[\binom{2k}{k}%
\binom{2N-2k}{N-k}\left( \frac{1}{2}\right) ^{2N}\bigg]^{1/2}  \notag \\
&&\hskip 0.5in\times \left\vert 2k\right\rangle _{u}\left\vert 2N-2k\right\rangle _{d}%
\text{,}\label{PN}
\end{eqnarray}%
where $\left\vert 2k\right\rangle _{u}\left\vert
2N-2k\right\rangle _{d}$ denote $2k$ photons and $2N-2k$ photons
in the up and down modes, respectively. This state has been
studied extensively in Refs.\cite{holland,campos}. The photon
number variance is
\begin{eqnarray}
\langle \Delta ^{2}N\rangle _{u} &=&\langle \hat a_{u}^{\dag }\hat
a_{u}\hat a_{u}^{\dag }\hat a_{u}\rangle -\langle \hat a_{u}^{\dag
}\hat a_{u}\rangle \langle \hat a_{u}^{\dag
}\hat a_{u}\rangle  \notag \\
&=&(N^{2}+N)/2\text{,}
\end{eqnarray}%
where $\langle \hat a_{u}^{\dag }\hat a_{u}\rangle =N$ and
$\langle \hat a_{u}^{\dag }\hat a_{u}\hat a_{u}^{\dag }\hat
a_{u}\rangle =(3N^{2}+N)/2$. From Eq.(1), we find the phase
fluctuation $\Delta \phi \thicksim \sqrt{2}/\sqrt{N^{2}+N}$. Thus
if we use this state to probe a small phase shift, the measurement
precision may approach $\sqrt{2}/N$ when $N$ is much larger than
$1$. This is close to the Heisenberg limit of $1/N$.

Let us now see how we can use the above state for phase
measurement. When the phase shift operator $\hat U=\exp (i\phi
\hat a_{u}^{\dag }\hat a_{u})$ acts on this state, the state
becomes:
\begin{eqnarray}
\left\vert \Psi _{2N}(\phi )\right\rangle
&=&\sum_{k=0}^{N}{(-1)^{N-k}\over 2^N}\bigg[\binom{2k}{k}\binom{2N-2k}{N-k}%
\bigg]^{1/2}  \notag \\
&&\hskip 0.5in \times e^{i2k\phi } \left\vert 2k\right\rangle _{u}\left\vert 2N-2k\right\rangle _{d}%
\text{.}\label{PNp}
\end{eqnarray}%
Refs.\cite{holland,campos} provided two methods to extract the
phase information: one is based on the variance of the photon
number difference \cite{holland} while the other on a parity
measurement \cite{campos}. However, none of them are easily
implemented experimentally. Here we present a different method for
phase measurement, based on the general guideline outlined in
Ref.\cite{ou2}. The idea is to compare the phase-shifted state in
Eq.(\ref{PNp}) with the original unshifted one in Eq.(\ref{PN}).
This can be done by making the projection of the state $\left\vert
\Psi _{2N}(\phi)\right\rangle$ onto the state $\left\vert \Psi
_{2N}\right\rangle$: $\langle\Psi_{2N}\left\vert \Psi
_{2N}(\phi)\right\rangle $, which is simply the measurement of the
projection operator $\hat{\mathbb{P}}\equiv \left\vert \Psi
_{2N}\right\rangle\langle \Psi _{2N}|$ on the state $\left\vert
\Psi _{2N}(\phi)\right\rangle$:
\begin{eqnarray}
P(\phi ) &=&\langle
\Psi_{2N}(\phi)|\hat{\mathbb{P}}|\Psi_{2N}(\phi)\rangle =
|\left\langle \Psi _{2N}\right\vert \Psi _{2N}(\phi )\rangle
|^{2}  \notag \\
&=&\Bigg\{\sum_{k=0}^{N}{\cos [\phi (2k-N)]\over
2^N}\binom{2k}{k}\binom{2N-2k}{N-k}\Bigg\}^{2}.~~~\label{Pp}
\end{eqnarray}

The projection operator $\hat{\mathbb{P}}\equiv \left\vert \Psi
_{2N}\right\rangle\langle \Psi _{2N}|$ can be constructed by the
method described in Ref.\cite{sun}. From the identity equation
\begin{eqnarray}
&&\sum_{k=0}^{N}(-1)^{N-k}x^{2k}y^{2(N-k)}\binom{N}{k} \left(
\frac{1}{2}\right) ^{2N}  \cr && = (x^2-y^2)^N/2^N =
[(x-y)/\sqrt{2}]^N[(x+y)\sqrt{2}]^N,~~~~~
\end{eqnarray}
we find that the projection $\hat {\mathbb{P}}$ is achieved by
measuring the probability of detecting $N$ photons at each outport
of a 50:50 beam splitter BS$_{2}$ (Fig.1b). The 2N-photon
coincidence rate is proportional to $P(\phi)$ in Eq.(\ref{Pp})

\begin{figure}[htb]
\begin{center}
\includegraphics[width= 3in]{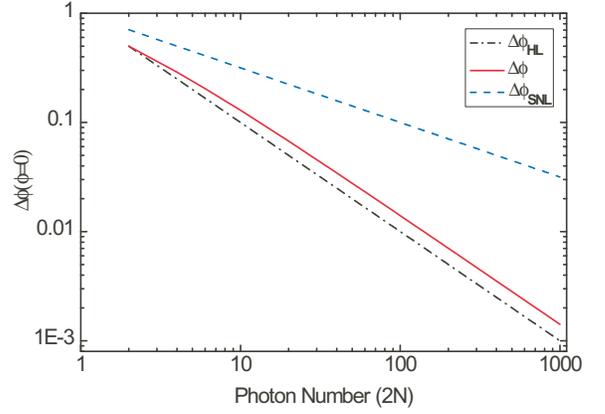}
\end{center}
\caption{Phase uncertainty versus photon number $N$. The solid
line shows the phase uncertainty at $\protect\phi =0$ using the
twin-photon state input and the projection measurement. The
dash-dotted line and dashed line show the
 Heisenberg limit $\Delta \protect\phi _{HL}$ and the standard quantum limit $%
\Delta \protect\phi _{SQL}$.}
\end{figure}

The phase uncertainty from the error propagation theory is given
by \cite{giovannetti}
\begin{equation}
\Delta \phi =\Delta P(\phi )\bigg/\bigg|\frac{\partial P(\phi
)}{\partial \phi }\bigg|,\label{dp}
\end{equation}
where $\Delta P(\phi)\equiv \langle \hat {\mathbb{P}}^2\rangle -
\langle \hat {\mathbb{P}}\rangle^2 = P(\phi)[1-P(\phi)]$. For the
case of $N=1$, the state $\left\vert \Psi _{2N}\right\rangle$ is
exactly the two-photon MES state and $P(\phi )=\cos ^{2}\phi $ so
that the phase uncertainty defined in Eq.(\ref{dp}) is $\Delta
\phi =1/2$. This is exactly the Heisenberg limit for two photons. With $N=2$, $P(\phi )=(\frac{3}{%
4}\cos 2\phi +\frac{1}{4})^{2}$, and $\Delta \phi
=1/\sqrt{12}=0.289$ in the limit $\phi \rightarrow 0$. This is
slightly larger than the Heisenberg limit of $\Delta \phi
_{HL}=1/4=0.25$, but is much less than the standard quantum limit
of $\Delta \phi _{SQL}=1/\sqrt{4}=0.5$ for four photons. A log-log
plot of the phase uncertainty derived from Eqs.(\ref{Pp},
\ref{dp}) with $\phi =0$ versus photon number $N$ is shown in
Fig.2 as the solid line together with the Heisenberg limit
(dash-dotted line) and the standard quantum limit (dashed line),
respectively. With a large photon number, the slope of the $\Delta
\phi $ is approaching $-1$, just as the slope of $\Delta \phi
_{HL}$. However, the plot of $\Delta \phi _{SQL}$ shows a slope of
$-1/2$. It is clear that $\Delta \phi$ is much better than $\Delta
\phi _{SQL}$. From the numerical calculation, we have $\Delta \phi
\sim 1.4\Delta \phi _{HL}$ for a large photon number.

\vskip 0.1in

\noindent {\it Experiment} -- Compared to the MES, the two-mode
twin Fock state and the state $\left\vert \Psi _{2N}\right\rangle
$ in Eq.(\ref{PN}) after the beam splitter can be prepared more
easily. For experimental implementation, two orthogonal
polarization modes are used instead of the spatial modes in Fig.1.
The polarization twin Fock state can be generated through the
process of type-II parametric down conversion (PDC). As shown in
Fig.3, a  BBO crystal cut for type-II beam-like parametric
down-conversion \cite{xiang} is pumped by 150 fs, 200mw
ultraviolet (390 nm) pulses to produce the photon pairs. The down
converted photons (780 nm) are coupled into the polarization
maintaining single-mode fiber. The two polarization modes are
combined by a polarization beam splitter (PBS$_{1}$) and
transmitted through a 3 nm interference filter (IF) and a half
wave plate (HWP$_{1}$).

\begin{figure}[tbh]
\begin{center}
\includegraphics[width= 3.3in]{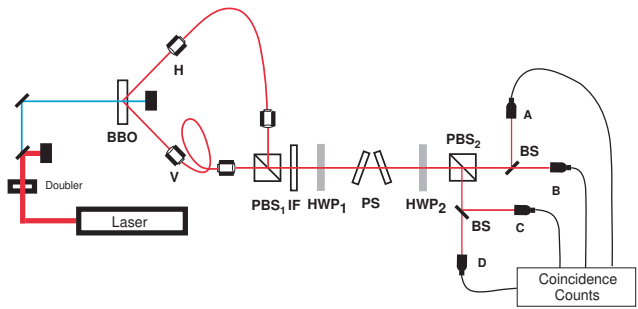}
\end{center}
\caption{(color online) Experimental setup. PBS: polarization beam
splitter; HWP: half wave plate; IF: interference filter; PS: phase
 shifter; BS: beam splitter.}
\end{figure}

It is easy to implement for photon states with small numbers such
as $N=1$, $2$, which can be realized with two and four detectors,
respectively (Fig.3). The phase shift operator $\hat U=\exp (i\phi
\hat a_{H}^{\dag }\hat a_{H})$ on horizontal polarization is
realized by changing the incident angles of the two identical
quartz crystals (PS), as illustrated in Fig.3. To avoid the shift
of the beam, the two quartz crystals are turned simultaneously in
opposite directions.
The projection measurement of $%
\left\vert \Psi _{2N}\right\rangle $ is achieved by 2N-fold
coincidence measurement after the half wave plate HWP$_{2}$ (set
at $22.5^{\circ }$) and PBS$_{2}$. The detection of $N$ photons in
each output port of PBS$_{2}$ indicates a successful projection to
$\left\vert \Psi _{2N}\right\rangle $. For two-photon case,
detectors A and C are used while all four detectors are involved
for four-photon case.
\begin{figure}[tbh]
\begin{center}
\includegraphics[width=2.5in]{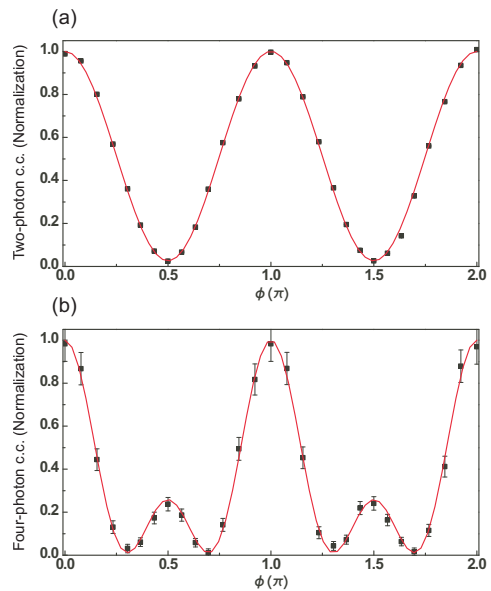}
\end{center}
\caption{(color online) Experiment data: (a) Two-photon
coincidence counts
between detectors A and C versus the single-photon phase difference $\protect%
\phi $. The continuous curve is a  least square fit to the function $%
P_{2}(\protect\phi )$. (b) Four-photon coincidence counts versus
$\protect\phi$. They are fitted to the function of
$P_{4}(\protect\phi )$. The data are normalized to the maximum
counts of $8837/s$ and $375/100s$ for (a) and (b), respectively.}
\end{figure}

Fig.4 shows the results of measurement with two-photon and
four-photon states. The data are normalized to the maximum count.
The result for two-photon state is illustrated in Fig.4(a). The
data are fitted to the function of $P_{2}(\phi)=(1+V\cos 2\phi
)/(1+V)$, with $V=(95.3\pm 0.1)\%$ as the visibility defined by
$V\equiv(C_{\max }-C_{\min })/(C_{\max
}+C_{\min })$. So the minimal phase uncertainty defined in Eq.(\ref{dp}) is $%
\Delta \phi =0.506\pm 0.001$ when $\phi =0$, whereas for an
average photon number of $2$, the standard quantum limit is
$\Delta \phi _{SQL}=0.707$.

Fig.4(b) shows the result of four-photon case. Since the four
photons are from two pairs of down-converted photons, we need to
consider the distinguishability between the two pairs
\cite{Ou99PRA}. Then the result is \cite{note}
\begin{equation}
P_{4}(\phi )={(1+2{\cal E/A})(3\cos 4\phi +4\cos 2\phi )+9+2{\cal
E/A}\over 16+16{\cal E/A}},
\end{equation}
where ${\cal E/A}$($\leq 1$) describes the degree of temporal
distinguishability between two pairs of photons. The closer to one
the value of ${\cal E/A}$ is, the more indistinguishable the pairs
are. In Fig.4(b), the solid curve is a least square fit to the
above function with ${\cal E/A}=0.93\pm 0.03$. This value
of ${\cal E/A}$ is consistent with that in previous study \cite{xiang}. From $P_{4}(\phi )$%
, we may derive the phase uncertainty from Eq.(\ref{dp}) and plot
it against the phase shift $\phi$, as shown in Fig.5. Compared
with the standard quantum limit (dashed), the phase uncertainty
from the four-photon state is better in the regions of $\phi \in
(k\pi -0.885,k\pi +0.885)$. The minimal phase uncertainty is
$\Delta \phi _{\min }=0.291\pm 0.001$ at $\phi = k\pi$.

\begin{figure}[tbh]
\begin{center}
\includegraphics[width= 3in]{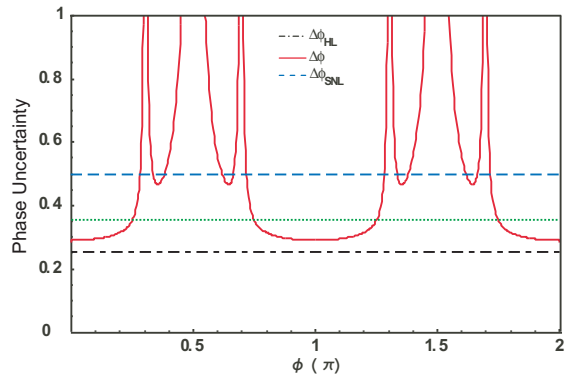}
\end{center}
\caption{(color online)Phase uncertainty versus phase shift. The
solid line shows the phase uncertainty using the four-photon
state. Correspondingly,
the dash-dotted line and dashed line show the cases of Heisenberg limit $%
\Delta \protect\phi _{HL}$ and standard quantum limit $\Delta \protect\phi _{SQL}$%
. }
\end{figure}

The improvement of precision comes from the multi-photon
interference effect due to multi-photon entanglement. The optimal
state is the MES. Other states with less entanglement, such as the
two-mode state in Eq.(\ref{PN}), may show a phase measurement
precision worse than the MES, but they may still give a precision
close to the Heisenberg limit. However, decoherence in the
multi-photon state will reduce the entanglement effect and
influence the phase measurement precision. For example, if the
four photons state from the parametric down conversion are
separated into two distinguishable pairs, i.e. ${\cal E/A}=0$
\cite{Ou99PRA}, the theoretical phase uncertainty is then $0.354$
(shown as the dotted line in Fig.5). The value of $0.354$ is still
better than the standard quantum limit $\Delta \protect\phi _{SQL}
= 0.5$. This is because there is still some partial entanglement
in this four-photon state (the down-converted two photons are
correlated and entangled).

In experiment, losses of photons and imperfection in detector
efficiency will decrease the success probability of projection
measurement. However, it will not affect the uncertainty of phase
measurement because the measurement fidelity remains unchanged.
The loss in projection measurement can be modelled as:
$\mathbb{M}=\eta ^{2}\left\vert \Psi _{2N}\right\rangle
\left\langle \Psi _{2N}\right\vert $, where the coefficient $\eta
$ depends on photon losses and detector efficiency. In the
calculation of $\Delta \phi $, we have $\Delta ^{2}M(\phi
)=\left\langle \Psi _{2N}(\phi )\right\vert
\mathbb{M}^{2}\left\vert \Psi _{2N}(\phi )\right\rangle
-(\left\langle \Psi _{2N}(\phi )\right\vert \mathbb{M}\left\vert
\Psi _{2N}(\phi )\right\rangle )^{2}=\eta ^{4}[P(\phi )-P^{2}(\phi
)]$ and $|\frac{\partial
M(\phi )}{\partial \phi }|=\eta ^{2}|\frac{\partial P(\phi )}{\partial \phi }%
|$. From Eq.(\ref{dp}), we find that the uncertainty of the phase
measurement is unchanged. The reason for this is that projection
measurement is a post-selection of the states with a fixed photon
number for the input state. Lost photons are excluded.  However,
the higher photon number states that may arrive from higher order
parametric down-conversion would contribute to the background and
should be subtracted. Otherwise it will decrease the fidelity of
projection measurement. Therefore, loss only decreases the success
probability and thus increases the number of trials required, that
is we will need $N/\eta$ photons instead of $N$ photons. But as
long as $\eta$ is fixed, the phase uncertainty will be $1.4/\eta
N\sim 1/N$, similar to the Heisenberg limit. Multi-photon
coincidence measurement usually has very low efficiency. It is
anticipated that as technology develops \cite{op}, new efficient
multi-photon detector will replace the low efficiency multi-photon
coincidence measurement for the projection measurement. Since our
experiment is a demonstration of principle, we set $\eta$ to 1 in
our analysis.

In conclusion, we discussed a new state projection measurement for
extracting the phase information with a twin Fock state, with a
for large photon number. The concept, i.e., a state preparation
for phase measurement and a projection for comparison, can be
generalized. Experimentally, we demonstrate the feasibility of the
scheme with a phase measurement precision of $0.506\pm 0.001$ for
two-photon state and $0.291\pm 0.001$ for four-photon state,
respectively. Unlike many other schemes realized so far
\cite{pan,mitchell,res,sun,Nagata}, our scheme can be applied to
large photon numbers.

After submitting our manuscript, we became aware of the work by
Nagata \textit{et al.} on a similar topic \cite{Nagata}.

\begin{acknowledgments}
This work was funded National Fundamental Research Program, the Innovation
funds from Chinese Academy of Sciences, National Natural Science Foundation
of China (Grant No.60121503). ZYO is also supported by the US National
Science Foundation under Grant No. 0427647.
\end{acknowledgments}


\begin{thebibliography}{99}
\bibitem{note} Because the two-photon interference has high visibility,
we do not consider photon distinguishability within one pair of
photons when we deduce $P_{4}(\phi )$.
\end{thebibliography}
\end{document}